# Evaluating the Performance of a Modified Skin Temperature Sensor for Lower Limb Prostheses: An Experimental Comparison*

Anirshu Devroy[1], Gregor Fritz[1] and Mathias Brandstötter[1]

*Abstract*— Current rehabilitation of lower limb prostheses has significant challenges, especially with skin conditions, irritation and discomfort. Understanding the skin temperature and having comfortable wearable sensors that would monitor skin temperature in a real-time outdoor environment would be useful. The system would help the user and orthopedic technician to provide feedback and changes that might be required in the prosthesis. Hence in this paper, a series of experiments are conducted in order to understand and characterize the system behavior and compare a general thermistor and a modified thermistor as a potential method of temperature measurement for outdoor usage of prostheses. The paper goes on to compare the different modified thermistors' behavior with their regular counterpart and highlights the challenges and improvement areas needed for such a modified thermistor for outdoor temperature monitoring in a prosthetic system. Initial results show that some of the modified thermistors showed better temperature recording compared to the rest. Finally, such modified thermistors can be a potential alternative for comfortable temperature measurement embedded in the prosthesis system. Such a system can provide valuable insights into temperature distribution and an early warning system for skin problems.

## I. INTRODUCTION

Amputation of a limb can have significant impacts on an individual's quality of life. With the growing population, the number of individuals affected by diabetes and other vascular conditions, such as peripheral artery disease, also increases. In the United States alone, it is estimated that 29.1 million people live with diabetes, which is one of the major causes of limb amputation [2]. Hence in order to improve the life quality of patients with limb loss, understanding the current challenges in rehabilitation procedures is important. In this paper, the focus will be on lower limb transfemoral amputees. Currently, a transfemoral prosthetic limb consists of a stump, a liner and a socket. The objective has always been to successfully integrate prostheses as part of the human skeleton without any pain or discomfort [14]. However, due to socket problems around 50% of limb amputees do not use their prostheses [13]. The usage of prostheses is even lower for people with transfemoral amputation [7]. Despite advancements in prosthetic technology, knowledge about the clinical state of the limb and comfort during prosthesis use remains limited. A major focus in the care of leg amputees is ensuring comfort and proper skin condition. Two of the primary challenges in this regard are discomfort associated with the socket of the prosthetic and skin problems [6]. The causes of the skin problems are due to various conditions such as ulcers, infections, etc. that the skin experiences during the use of the prosthesis [11]. Therefore, temperature and humidity sensors are needed to obtain data to quantify and support the patient's perceptions. The information on temperature and moisture is crucial, as a wrong ratio could cause infection and skin damage, sometimes even without causing the patient noticeable discomfort [9]. Hence, in order to better monitor the skin condition, the usage of temperature sensors and humidity sensors are needed [4]. Day-to-day mobility activities and post-exercise could consequent in the temperature of the skin not coming back to its normal state [15]. Currently, there is no established system known to the author that collects real-time-temperature data from the prostheses system during outdoor activities. Such a system would contribute to an understanding of the thermal map of the limb prostheses in terms of activities [1]. This would help technicians and doctors to better predict and improve the comfort of prosthetic systems. However, the general usage of thermistors is uncomfortable and can only be used for short-duration temperature sensing [12]. Discomfort can be seen as a qualitative term for a prosthesis. In this paper the aspect of discomfort was broken down into three basic parts:

1) Discomfort due to leg movement
2) Discomfort due to sensor contact
3) Discomfort due to temperature changes

The focus of this paper is on the last two problems. Firstly, the temperature sensors should be comfortable and, in the future, also portable. This could potentially understand the skin conditions and help in better predictions of skin problems. Hence, the objective is to use a thermistor that would be comfortable and would provide real-time data. As it was mentioned earlier, thermistors are not comfortable, therefore the thermistor was modified by covering it with a silicone layer such that it is pleasant to wear for a long duration. Then the paper goes further to understand and verify whether the modified thermistor is behaving like its regular counterpart and if there is any significant loss of data. The experiments were conducted using all sensors with skin contact to make sure there are no major properties or behavior changes. Further, a description of the observed data among the sensors for various case studies and the behavior of the sensors are illustrated. In general, the skin temperature rapidly rises due to any sudden exercise movements [8]. Hence, the paper

*The research leading to these results has received funding from the Federal Ministry for Digital and Economic Affairs (BMDW) within the framework of Bridge, 34st call of the Austrian Research Promotion Agency (FFG) - project number 891132 (AMASE).

[1]All authors are with ADMiRE Research Center – Additive Manufacturing, Intelligent Robotics, Sensors and Engineering, Carinthia University of Applied Sciences, Europastrasse 4, 9524 Villach, Austria
{a.devroy,g.fritz,m.brandstoetter}@cuas.at

compares the sensor reading based on various exercises. Finally, the benefits and further research are highlighted.

## II. STATE OF THE ART

The idea of using wireless temperature sensors is already part of a study. In order to measure real-time data of the socket and liner interface, a pyroelectric effect on a PZT(Lead Zincronet Titanet) device was used. However, the results from this sensor had a low resolution[5]. In another experiment, a flexible temperature sensor was developed which was integrated into the yarns of the textile. The experimental setup was realized as an armband. However, there was a reduction in effective sensitivity due to muscle contraction[10]. In 2018 a temperature sensor along with a strain sensor was developed using the concept of polarization maintaining fiber and multi-mode fiber. However, the research had no testing with actual humans and had no results related to the sensitivity of the temperature sensor during activities [16]. One of the first wireless temperature sensors were tested in 2007. NTC(Negative Temperature Coefficient) temperature sensors were tested in this study. However, this work was on upper limb prosthesis using the temperature sensor as feedback[3].

## III. METHOD AND PATIENT

In order to measure the temperature of the skin PT100 sensors were selected. The PT100 sensor is very small in size and can be easily fitted into the liner of the prosthesis for continuous measurement. The sensor itself is (2x4x0.5) mm. It was assumed that the resistance of the sensor is linearly related to the temperature. There were 6 sensors of different thicknesses prepared for the experiment. The sensors were color coded to understand the thickness of each of them. In this article, the color-coding nomenclature will be followed throughout the rest of the article. The color coding is Black, Blue, Red, Yellow, and White with the increasing order of silicone thickness layer, as can be seen in Tab. I.

| Sensor | Thickness | Sensor | Thickness |
|---|---|---|---|
| 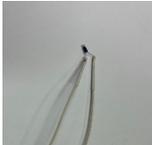 | No silicone 0 mm | 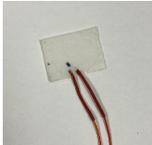 | Red 0.63 mm |
| 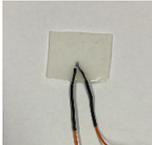 | Black 0.21 mm | 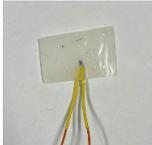 | Yellow 0.84 mm |
| 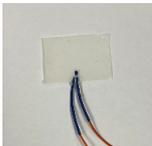 | Blue 0.42 mm | 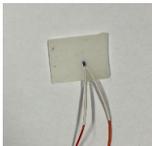 | White 1.05 mm |

TABLE I: Sensor designation and thickness

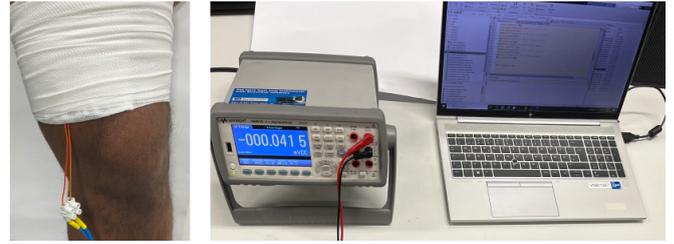

Fig. 1: Setup to measure the limp temperature and to process the data

Lastly, there is one PT100 sensor without a silicone layer, it is a normal PT100 sensor for reference. The silicone used was Dragonskin-10-fast. The order of thickness for each silicone was 0.21 mm for the black one and increased up to 1.05 mm. Each color in the order had one extra layer of 0.21 mm thickness placed. The thickness of silicone here referred to is the layer between the PT100 temperature sensor and the skin they are attached to. Each of the sensors was placed on different positions of a healthy limb to evaluate the sensitivity of the sensor. For each position on the limb, 3 sets of experiments were conducted, once for each sensor. The experiments were 1) Walking 2) Squats 3) Running. However, to better understand the sensitivity and behavior of the sensors another pair of experiments was conducted with each sensor 1) when they are not in contact and 2) when they are in contact by a pinch grip. The position of the sensors for the current set of experiments was above the knee. The position selection of the different sensors was selected based on this article [9]. The position-based different experiment was conducted on a person with a healthy limb without any amputation. A single position separate set of the same experiments was conducted for two other people of 30 years and 35 years to evaluate the temperature value compared to the first participant. The experimental setup is shown in Fig. 1 and illustrates the sensor attached to the patient as well as the measurement device.

The readings of the resistance were collected from a Keysight digital multimeter 34461A. The multimeter was connected via USB to the desktop and MATLAB was used to read the signals from the digital multimeter using the MATLAB instrumentation control toolbox. The experiment was conducted for 100 seconds. The data collection rate was 1 value per second. The experimental layout is visualized in Fig. 2. It was also considered that silicone is not a good conductor of heat, hence, there might be less sensitivity of the thermistor as the thickness increases of the thermistors compared to the regular PT100 sensor. Therefore, the experiment was started with the least possible layer of silicone, 0.21mm, and kept on increasing to understand how much the increase in thickness affects the sensitivity compared to its regular counterpart.

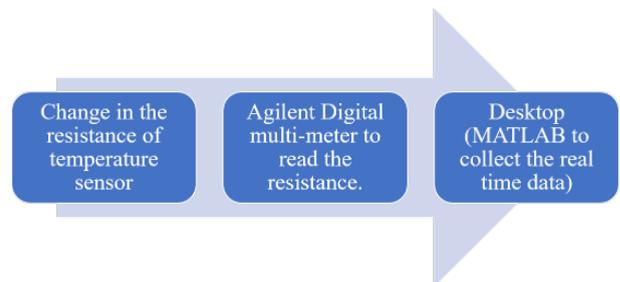

Fig. 2: The connection between sensor and limb

## IV. RESULT AND DISCUSSION

The results of the different sensors based on different positions were obtained to evaluate the temperature changes, the sensor behavior, and how the sensitivity of the sensors changes based on the increase in the thickness of silicone. It should be noted that the experiments were performed continuously one after the other, consequently, the average temperature range is higher in later experiments. Also, the exercise movements, the muscle contractions, and the leg movements made significantly influenced the temperature behavior and its signal. The results are classified based on the different positions. The position selection was based on the article [9] The sensor placement positions are as follows:

1) Position 1 as T1
2) Position 2 as T2
3) Position 3 as T7′
4) Position 4 as T8′
5) Position 5 as T3

The position of the various sensor positions is outlined in Fig. 3. Here, the transfemoral (TFA) amputation is considered, as it is shown on the left-hand side of said figure. The TTA on the right-hand side stands for the transtibial amputee. The obtained results were categorized into different cases and discussed in the following 4 sections.

### A. CASE 1: Sensor testing

In this section, the thermistors were compared with and without skin contact, while temperature variations were recorded. The temperature value for the sensors with no contact is shown in Fig 4.

In Fig. ?? it can be seen that there is a steady decline in temperature. A possible explanation would be that the room temperature, in which the temperature was recorded, was very low. Consequently, after switching on the sensors it needs time to adjust. Reaching the temperature reading of the room could take some time

A separate experiment was conducted where the sensors were pinched. The setup of the experiment is shown in Fig. 5. As can be seen, the index finger and the thumb were used to grip the temperature sensor. A pinch contact was established for the first 50 seconds and then released, followed by another pinch contact after approximately 80 seconds. This experiment helps to understand how long the sensors take to adjust and how much the slope of the graph changes for each of the stages' temperatures. The results obtained are shown in Fig. 6.

As can be seen from Fig. 6 as the layer thickness of silicone increases, the sensitivity of the sensor decreases. The orange line with no silicone has the highest sensitivity. The value of the temperature shoots up when the pinch contact is made. Followed by when the contact is released there is a free fall in the temperature value and it rises as soon as the contact is made again. It is to be observed that the temperature follows a hyperbolic curve when returning to its steady state. As the thickness of the silicone increases the time taken to reach the steady state takes longer before and after contact with the sensors. The white temperature sensor representing the blue line has the least changes in temperature. This signifies that the white sensor (the thickest sensor) has the least sensitivity of all sensors tested.

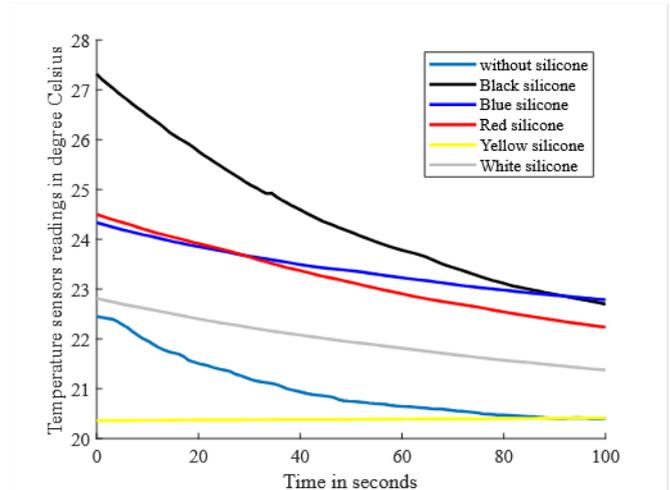

Fig. 4: No contact temperature reading

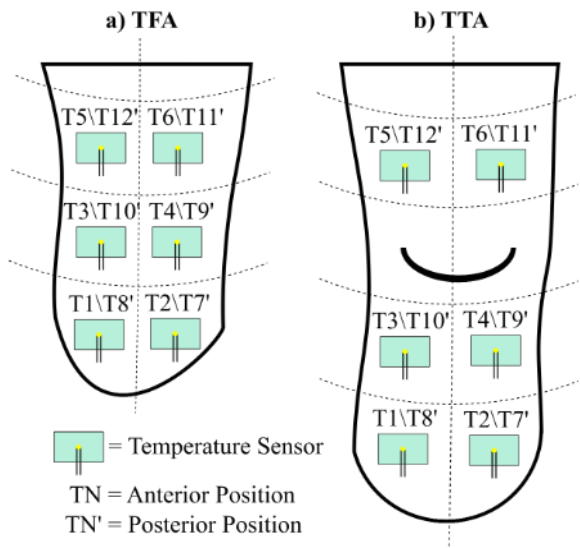

Fig. 3: The position of the sensor [15]

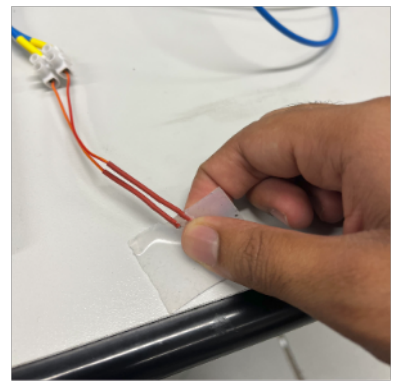

Fig. 5: Pinch grip of the temperature sensor

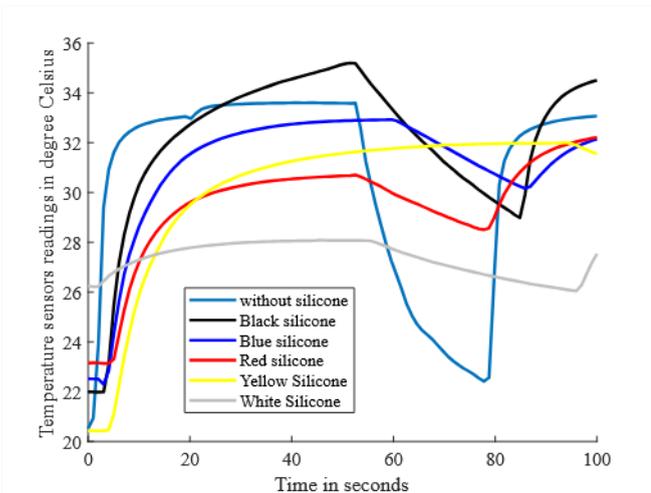

Fig. 6: Pinch contact temperature readings

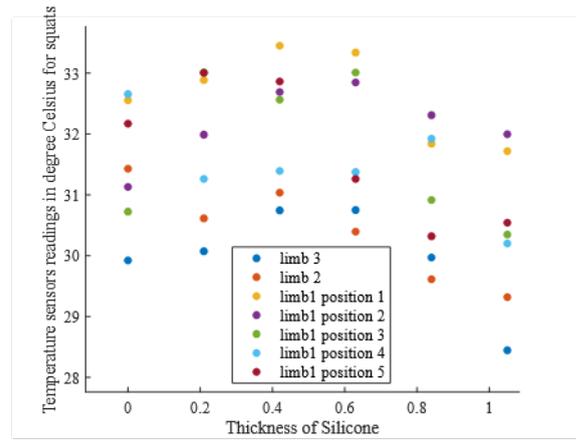

Fig. 8: Squat – temperature readings

*B. CASE 2: Walking Exercise*

In this case, a median value was calculated for the different positions and thickness for a temperature reading duration of 100 seconds. The median value was chosen to better represent the saturation temperature and ignore outlier values. The 3 patients were made to walk, during the temperature recording session. For one of the patients, the position variance-based temperature recording was also conducted. Limb 2 and limb 3 are participant 2 and 3. As can be seen from Fig. 7 the sensor thickness seems to significantly affect the median value of the temperature data. When the thickness of the silicone increases the temperature sensor shows a lower temperature recording overall. However, it should also be noted that there are not many sensitivity changes due to silicone thickness for the initial thin layer of thickness. Fig. 7 also illustrates that in some cases the temperature is measuring higher values compared to the corresponding no silicone-based temperature sensor. There seems to be a rise in the temperature recording value for the thin layers of the temperature i.e blue, red, and black. For all the walking exercises participants had the liberty in choosing the walking speed on the floor and no treadmill or any other devices were used.

*C. CASE 3: Squat Exercise*

A similar experimentation as in the previous case was replicated but with squatting exercises. The results obtained also show a very similar trend as the walking scenario in that the sensors with silicone layers from 0.21-0.63 seems to record a very higher overall range of temperature compared to the non-silicone counterpart. The resulting graph is shown in Fig. 8. Also in comparison to the previous graph of Fig. 7 the maximum value of the temperature reached is higher in Fig. 8 compared to walking since squatting is a more intense exercise than walking. It is also to be observed that the changes in the position of the sensor can result in significant changes in the value of the skin temperature. It should be noted that there was no frequency in the squatting exercises and was done as a stationary exercise and the participants were asked to make as many squats as possible in 100 seconds.

*D. CASE 4: Running Exercise*

The results of the running exercise are shown in Fig. 9. In the running exercise, the highest temperature values of all exercises were recorded. Hence, from observation, the white and the yellow sensors having a thick layer seem to read a lower temperature value compared to the sensors with thinner silicon layers. Having such a diverse set of experiments, the

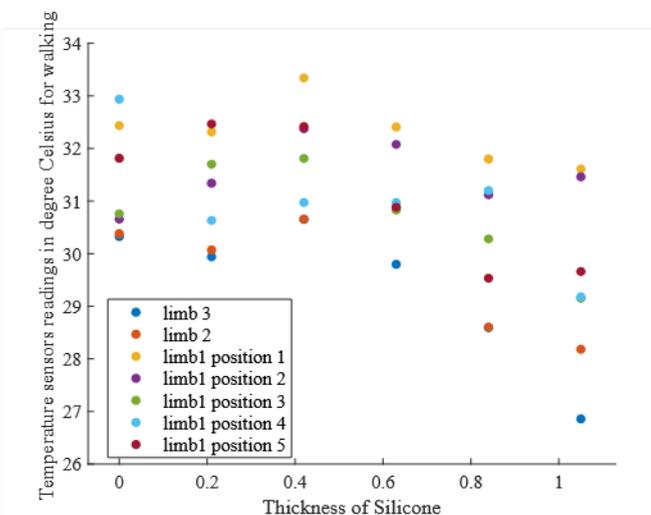

Fig. 7: Walking – temperature readings

119

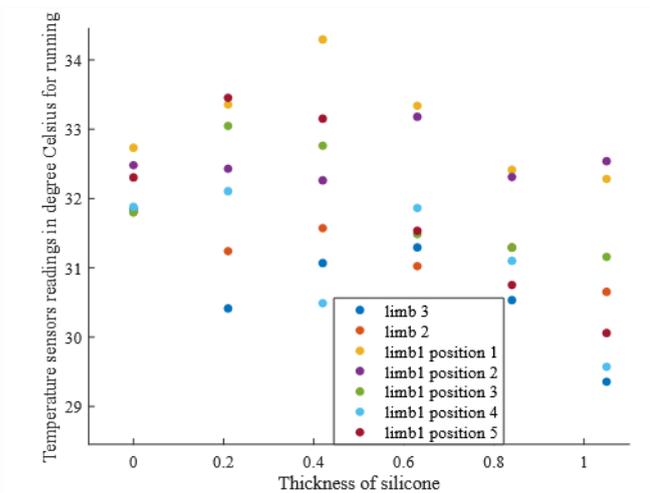

Fig. 9: Running – temperature readings

possible explanation for the low values of the white and yellow sensors would be that the thick silicone layer acts as an insulator for every case. Hence, using such temperature sensors may lead to data inaccuracies since it is managing to record lower than the possible actual value. Also, the other black, blue and red seem to be the outliers in all of the experiments. A number of reasons can be the cause of such behavior, like trapping of the body heat, and thin thickness layer thickness contributing to such an effect. Overall all of the exercise cases seem to have common traits of lower values for the yellow and white sensor and outlier values for the black, blue, and red. The running exercise was done on the floor without any external devices where the participants had liberty of choosing the speed.

## V. CONCLUSION AND FUTURE WORKS

After thorough experimentation it was observed that:

1) The repeated experimentation resulted in damage to the sensors. Hence robustness is a question that would have to be addressed in the future. However, the thickest sensor, which is the white sensor, survived the entire series of experiments without any replication. Hence to make the sensor more robust the reading has to be compensated. An optimized sensor thickness vs robustness is something to look for in thread.
2) From cases 2 to 4 it can be observed that the blue, black, and red sensors have higher temperature recordings than anticipated. The blue sensor seems to record a range even higher compared to the black and the red ones in general. The proper reason is yet to be known.
3) To understand the outlier behavior of such a sensor a more loom experiment is to be conducted to understand how much these temperature changes along with a detailed note of what kind of limb movements the patient is making. The changes in the temperature data are very small hence the readout devices resolution needs to be very high. Otherwise, there would be more inaccuracies.
4) The experiments also need to be performed with a real prosthesis wearer to understand how the sensor behaves with an amputated leg as opposed to a healthy leg.
5) Based on the initial results if the requirement is for long-term robustness thick layer will work however there would be a low sensitivity to changes in the temperature. Hence on of the key challenges is to determine the sensitivity vs robustness However, for the short term, a user may use one of the thin sensors like black, blue, or red however more detailed results are needed to derive a conclusion about which would be the best suited for short terms needs
6) In terms of comfort the thinnest silicone layer manages to do the job based on the 3 participants experience. In the future focus lies on evaluating the sensor behavior in more broad cases in the understanding of resolution requirements, duration of measurement, and application.

In the future, we will focus on the aforementioned points detailing enhancements to the sensor, with the goal of improving its performance in accordance with the intended application.

## ACKNOWLEDGMENT

We would like to thank Marius Laux and Sebastian Spinzky for their support in the conducting of the experiments in the lab.

## REFERENCES

[1] S. A. Abbood, Z. Wu, and B. Sundén, "Thermal assessment for prostheses: state-of-the-art review," *International Journal of Computational Methods and Experimental Measurements*, vol. 5, no. 1, pp. 1–12, 2017.
[2] J. Bennett, "Limb loss: The unspoken psychological aspect," *Journal of vascular nursing*, vol. 34, no. 4, pp. 128–130, 2016.
[3] Y. Cho, K. Liang, F. Folowosele, B. Miller, and N. V. Thakor, "Wireless temperature sensing cosmesis for prosthesis," in *2007 IEEE 10th International Conference on Rehabilitation Robotics*. IEEE, 2007, pp. 672–677.
[4] G. Ciuti, L. Ricotti, A. Menciassi, and P. Dario, "Mems sensor technologies for human centred applications in healthcare, physical activities, safety and environmental sensing: A review on research activities in italy," *Sensors*, vol. 15, no. 3, pp. 6441–6468, 2015.
[5] A. Davidson, A. Buis, and I. Glesk, "Toward novel wearable pyroelectric temperature sensor for medical applications," *IEEE Sensors Journal*, vol. 17, no. 20, pp. 6682–6689, 2017.
[6] N. L. Dudek, M. B. Marks, S. C. Marshall, and J. P. Chardon, "Dermatologic conditions associated with use of a lower-extremity prosthesis," *Archives of physical medicine and rehabilitation*, vol. 86, no. 4, pp. 659–663, 2005.
[7] D. Durmus, I. Safaz, E. Adıgüzel, A. Uran, G. Sarısoy, A. S. Goktepe, and A. K. Tan, "The relationship between prosthesis use, phantom pain and psychiatric symptoms in male traumatic limb amputees," *Comprehensive Psychiatry*, vol. 59, pp. 45–53, 2015.
[8] G. K. Klute, E. Huff, and W. R. Ledoux, "Does activity affect residual limb skin temperatures?" *Clinical Orthopaedics and Related Research®*, vol. 472, no. 10, pp. 3062–3067, 2014.
[9] R. Kumar, F. Kimura, K. W. Ahn, Z.-H. Hu, Y. Kuwatsuka, J. P. Klein, M. Pasquini, K. Miyamura, K. Kato, A. Yoshimi, *et al.*, "Comparing outcomes with bone marrow or peripheral blood stem cells as graft source for matched sibling transplants in severe aplastic anemia across different economic regions," *Biology of Blood and Marrow Transplantation*, vol. 22, no. 5, pp. 932–940, 2016.

[10] P. Lugoda, J. C. Costa, C. Oliveira, L. A. Garcia-Garcia, S. D. Wickramasinghe, A. Pouryazdan, D. Roggen, T. Dias, and N. Münzenrieder, "Flexible temperature sensor integration into e-textiles using different industrial yarn fabrication processes," *Sensors*, vol. 20, no. 1, p. 73, 2019.

[11] H. E. Meulenbelt, J. H. Geertzen, P. U. Dijkstra, and M. F. Jonkman, "Skin problems in lower limb amputees: an overview by case reports," *Journal of the European Academy of Dermatology and Venereology*, vol. 21, no. 2, pp. 147–155, 2007.

[12] L. Paternò, M. Ibrahimi, E. Gruppioni, A. Menciassi, and L. Ricotti, "Sockets for limb prostheses: a review of existing technologies and open challenges," *IEEE Transactions on Biomedical Engineering*, vol. 65, no. 9, pp. 1996–2010, 2018.

[13] G. E. Reiber, L. V. McFarland, S. Hubbard, C. Maynard, D. K. Blough, J. M. Gambel, and D. G. Smith, "Servicemembers and veterans with major traumatic limb loss from vietnam war and oif/oef conflicts: Survey methods, participants, and summary findings," *Journal of rehabilitation research and development*, vol. 47, no. 4, pp. 275–298, 2010.

[14] R. Safari, "Lower limb prosthetic interfaces: Clinical and technological advancement and potential future direction," *Prosthetics and Orthotics International*, vol. 44, no. 6, pp. 384–401, 2020.

[15] R. J. Williams, A. Takashima, T. Ogata, and C. Holloway, "A pilot study towards long-term thermal comfort research for lower-limb prosthesis wearers," *Prosthetics and Orthotics International*, vol. 43, no. 1, pp. 47–54, 2019.

[16] R. Xing, C. Dong, Z. Wang, Y. Wu, Y. Yang, and S. Jian, "Simultaneous strain and temperature sensor based on polarization maintaining fiber and multimode fiber," *Optics & Laser Technology*, vol. 102, pp. 17–21, 2018.